\documentclass[prd,aps,nofootinbib,superscriptaddress,floatfix,notitlepage]{revtex4-1}
\pdfoutput=1

\usepackage{epsfig}
\usepackage{amsmath}

\input{epsf}

\begin{document}

\hfill CERN-TH-2020-133
\vspace{0.5cm}

\title{From Ji to Jaffe-Manohar orbital angular momentum
in Lattice QCD using a direct derivative method}

\author{M.~Engelhardt}
\email{engel@nmsu.edu}
\affiliation{Department of Physics, New Mexico State University,
Las Cruces, NM 88003, USA}

\author{J.~R.~Green}
\affiliation{Theoretical Physics Department, CERN, 1211 Geneva 23,
Switzerland}

\author{N.~Hasan}
\affiliation{Bergische Universit\"at Wuppertal, 42119 Wuppertal, Germany}

\author{S.~Krieg}
\affiliation{Bergische Universit\"at Wuppertal, 42119 Wuppertal, Germany}
\affiliation{IAS, J\"ulich Supercomputing Centre, Forschungszentrum 
J\"ulich, 52425 J\"ulich, Germany}

\author{S.~Meinel}
\affiliation{Department of Physics, University of Arizona, Tucson,
AZ 85721, USA}

\author{J.~Negele}
\affiliation{Center for Theoretical Physics, Massachusetts Institute of
Technology, Cambridge, MA 02139, USA}

\author{A.~Pochinsky}
\affiliation{Center for Theoretical Physics, Massachusetts Institute of
Technology, Cambridge, MA 02139, USA}

\author{S.~Syritsyn}
\affiliation{RIKEN BNL Research Center, Brookhaven National Laboratory,
Upton, NY 11973, USA}
\affiliation{Department of Physics and Astronomy, Stony Brook University,
Stony Brook, NY 11794, USA
\vspace{0.5cm}
}

\begin{abstract}
A Lattice QCD approach to quark orbital angular momentum in the proton
based on generalized transverse momentum-dependent parton distributions
(GTMDs) is enhanced methodologically by incorporating a direct derivative
technique. This improvement removes a significant numerical bias that
had been seen to afflict results of a previous study. In particular, the
value obtained for Ji quark orbital angular momentum is reconciled with
the one obtained independently via Ji's sum rule, validating the GMTD
approach. Since GTMDs simultaneously contain information about the quark
impact parameter and transverse momentum, they permit a direct evaluation
of the cross product of the latter. They are defined through proton
matrix elements of a quark bilocal operator containing a Wilson line;
the choice in Wilson line path allows one to continuously interpolate
from Ji to Jaffe-Manohar quark orbital angular momentum. The latter is
seen to be significantly enhanced in magnitude compared to Ji
quark orbital angular momentum, confirming previous results.
\end{abstract}

\maketitle

\section{Introduction}
The manner in which the spin of the proton arises from the spins and
orbital angular momenta of its quark and gluon constituents has been
the object of sustained study. Efforts to resolve this so-called
proton spin puzzle were sparked by the finding, in EMC experiments
\cite{emc1,emc2}, that the quark spins alone fail to provide a
satisfactory account of the proton's overall spin. Naturally, also the
methods of Lattice QCD have been brought to bear on the problem, with the
standard calculational scheme relying on Ji's sum rule \cite{jidecomp}.
The sum rule relates the total quark angular momentum $J$ to specific
moments of generalized parton distributions (GPDs), and by combining this
with a calculation of the quark spin $S$ \cite{lanl,liu2018}, one
can then also isolate the quark orbital angular momentum $L=J-S$
\cite{LHPC_1,LHPC_2,reg2004,reg2019,liu2015,etm2017,etm2020}.
The more recent studies have furthermore begun to gain control
over the gluon angular momentum \cite{liu2015,etm2017,etm2020} and
gluon spin \cite{liu2017} contributions.

The aforementioned indirect approach to quark orbital angular momentum is
limited specifically to the Ji decomposition of proton spin associated
with Ji's sum rule. However, the definition of quark orbital angular
momentum in QCD is not unique, since the matter degrees of freedom
in a gauge theory cannot be unambiguously separated from the gauge
degrees of freedom. Quarks necessarily carry gauge fields along with
them, and it is a matter of definition to what extent these are
included in the evaluation of quark orbital angular momentum. In
addition to the Ji decomposition of proton spin, another widely studied
decomposition scheme is the one due to Jaffe and Manohar \cite{jmdecomp}.
It possesses the conceptual advantage of allowing for a partonic
interpretation of the angular momentum distributions.

A formulation that offers a direct path to evaluating quark orbital angular
momentum and that encompasses both of the aforementioned decompositions
is the one in terms of generalized transverse momentum-dependent parton
distributions (GTMDs) \cite{mms,lorce,leader}. GTMDs, as functions of quark
transverse momentum $k_T $ as well as momentum transfer $\Delta_{T} $,
are related, through the Fourier conjugate pair $(\Delta_{T} ,b_T )$
(with $b_T $ denoting the quark impact parameter), to Wigner functions
that allow one to sample directly orbital angular momentum $b_T \times k_T $.
GTMDs are defined through a quark bilocal operator containing a Wilson line,
and it is the choice of path of the Wilson line that allows one to access
different definitions of quark orbital angular momentum. In particular,
it was realized in \cite{hatta} that Jaffe-Manohar orbital angular momentum
is associated with a staple-shaped path in the limit of infinite staple
length, an operator type extensively studied in the context of standard
transverse momentum-dependent parton distributions (TMDs). On the other
hand, Ji orbital angular momentum results from using a straight path
\cite{jist,burk,eomlir,eomlong}.

An initial Lattice QCD exploration of the GTMD approach to quark orbital
angular momentum was undertaken in \cite{jitojm}. By varying the staple
length of a staple-shaped Wilson line path in small steps, a
quasi-continuous, gauge-invariant interpolation between the Ji and
Jaffe-Manohar limits was realized. In performing this study, it was
possible to take recourse to concepts and methods from standard
Lattice TMD studies \cite{straightlett,straightlinks,tmdlat,bmlat,rentmd},
since, as noted above, the same operator structure enters; GTMD matrix
elements only differ by their additional dependence on the momentum
transfer $\Delta_{T} $. Jaffe-Manohar quark orbital angular momentum
was seen to be significantly enhanced in magnitude relative to its Ji
counterpart. The results obtained in \cite{jitojm} were, however,
affected by one substantial shortcoming: Although the relative comparison
between Ji and Jaffe-Manohar quark orbital angular momentum could be
expected to be trustworthy, in absolute terms, the orbital angular
momenta were significantly underestimated for a technical reason.
Namely, the weighting by $b_T $ in orbital angular momentum
$b_T \times k_T $ corresponds to computing a derivative with respect to
$\Delta_{T} $ of the relevant GTMD matrix element. This derivative was
realized via a finite difference over a momentum interval that was much
too large to yield an accurate estimate. This became directly apparent
in comparing the result for Ji quark orbital angular momentum with the
corresponding result obtained using Ji's sum rule. The discrepancy
amounted to approximately a factor 2. The present work resolves
this discrepancy by incorporating a direct derivative
method to evaluate the $\Delta_{T} $-derivative. This methodological
improvement removes the described bias by construction, and will be
seen to reconcile the results obtained through the GTMD approach and
through Ji's sum rule. This validates the GTMD approach as implemented
here.

The present study also uses pion mass $m_{\pi } =317\, \mbox{MeV} $,
which is significantly lower than the mass $m_{\pi } =518\, \mbox{MeV} $
used in the initial exploration \cite{jitojm}.

\section{Generalized transverse momentum-dependent (GTMD) approach to quark
orbital angular momentum}
\label{oamsec}
As laid out in detail in \cite{jitojm}, cf.~also \cite{leader,lorce},
the longitudinal quark orbital angular
momentum component $L_3 $ in a longitudinally polarized proton propagating
in the $3$-direction with momentum $P$ can be evaluated within Lattice QCD
in units of the number of valence quarks $n$ via
\begin{equation}
\frac{L_3 }{n} = \frac{1}{a} \epsilon_{ij}
\left. \frac{\frac{\partial }{\partial \Delta_{T,j} }
\left( \Phi (a\vec{e}_{i} ) - \Phi (-a\vec{e}_{i} ) \right) }{
\Phi (a\vec{e}_{i} ) + \Phi (-a\vec{e}_{i} )}
\right|_{\Delta_{T} =0}
\label{ratiodef}
\end{equation}
(summation over $i,j$ implied), with the proton matrix element
\begin{equation}
\Phi (z_T) = \langle P+\Delta_{T}/2 , S=\vec{e}_{3} |
\overline{\psi}(-z_T /2) \gamma^+ U[-z_T /2,z_T /2]
\psi(z_T /2) | P-\Delta_{T}/2 , S =\vec{e}_{3} \rangle \ .
\label{medef}
\end{equation}
Here, $\vec{e}_{3} $ denotes the unit vector in longitudinal direction,
whereas $\vec{e}_{i} $ is a unit vector in a transverse direction; $a$
denotes the lattice spacing. The momentum transfer $\Delta_{T} $ and
the operator separation $z_T $ are purely transverse and orthogonal
to each other. $U$ is a Wilson line connecting the quark operators
$\overline{\psi} $, $\psi $; its path remains to be specified. This
structure can be understood as follows: In the limit $z_T \rightarrow 0$,
the operator in $\Phi (z_T) $ reduces to the light-cone $+$-component of
the vector current, and therefore, at $\Delta_{T} =0$, $\Phi (z_T) $
simply counts valence quarks (up to a normalization factor $2P^{+} $). This
motivates the denominator in (\ref{ratiodef}). Note the use of the nonlocal
current with separation $a$, matching the numerator; this will be
revisited below. Consider now taking the $\Delta_{T} $-derivative of
$\Phi (z_T) $, and evaluating at $\Delta_{T} =0$. The momentum transfer
$\Delta_{T} $ is Fourier conjugate to the quark impact parameter $b_T $,
and therefore, this operation amounts to weighting the counting of quarks by
their impact parameter $b_T $. Likewise, the operator separation $z_T $ is
Fourier conjugate to the quark transverse momentum $k_T $. Thus, taking
the derivative with respect to $z_T $ and evaluating at $z_T =0$ amounts
to weighting the counting of quarks by their transverse momentum $k_T $.
The numerator in (\ref{ratiodef}), together with the division by $a$, is the
appropriate finite-difference realization of such a $z_T $-derivative;
at finite lattice cutoff $a$, distances smaller than $a$ cannot be
resolved. In aggregate, therefore, (\ref{ratiodef}) evaluates the total
$L_3 = b_T \times k_T $ of the quarks in the proton, normalized to the
number of valence quarks $n$.

\begin{figure}[h]
\centerline{\psfig{file=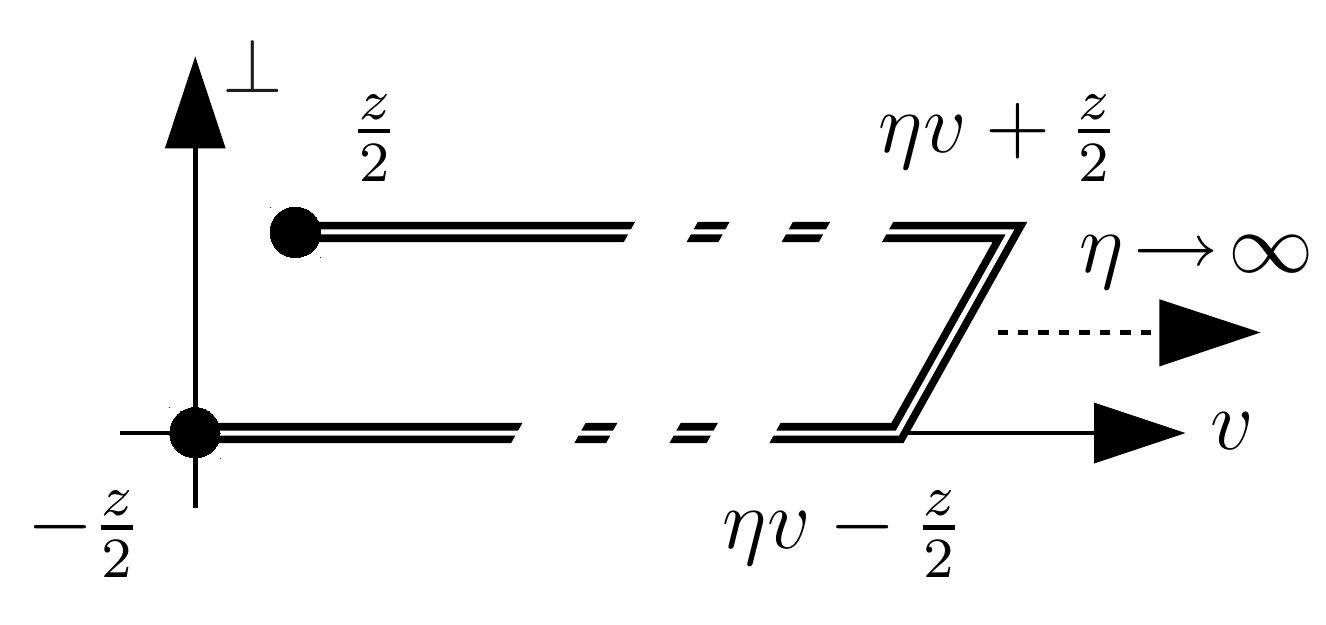,width=7cm} }
\caption{Path of the gauge connection $U$, cf.~(\ref{udef}), in the
correlator (\ref{medef}).}
\label{staplefig}
\end{figure}

An important role in the evaluation of quark orbital angular momentum
falls to the Wilson line $U$ in (\ref{medef}). In a gauge theory, the
matter degrees of freedom cannot be treated in complete isolation; they
necessarily carry gauge fields with them, and the evaluation of quark
orbital angular momentum depends on a definition of what part of the
overall gauge field to apportion to quarks as one decomposes
orbital angular momentum into a quark and a gluon part. The path of
the Wilson line $U$ carries this information. In the following,
staple-shaped paths
\begin{equation}
U\equiv U[-z/2,\eta v-z/2,\eta v+z/2,z/2]
\label{udef}
\end{equation}
will be considered, in which the points listed in the argument of $U$ are
connected by straight Wilson lines, as illustrated in Fig.~\ref{staplefig}.
The direction of the staple legs is defined by the vector $v$, with their
length scaled by the parameter $\eta $. For $\eta =0$, the path reduces
to a straight line between $-z/2$ and $z/2$. For ease of notation, in
the following, $\eta $ will also be allowed to be negative to reverse the
direction of the staple (keeping $v$ fixed). This class of gauge links
contains two important limits: $\eta =0$ corresponds to the Ji decomposition
of proton spin \cite{jist,burk,eomlir,eomlong}, whereas
$\eta \rightarrow \pm \infty $, with $v$ pointing in a light-like
direction, corresponds to the Jaffe-Manohar decomposition of proton
spin \cite{hatta,burk,eomlong}. By varying $\eta $ continuously, a
gauge-invariant interpolation between these two decompositions can be
obtained.

The operator in (\ref{medef}), extracting information about quark momentum
from the proton state, is of the standard form used in the definition of
transverse momentum-dependent parton distributions (TMDs)
\cite{collbook,ji04,aybat}. The matrix element (\ref{medef})
only differs from the standard TMD correlator by the introduction of
the nonvanishing momentum transfer $\Delta_{T} $ in the external states,
defining a generalized TMD (GTMD) correlator \cite{mms} in which the quark
momentum information is supplemented by quark impact parameter information.
Consequently, considerations from the standard TMD framework can be applied
\cite{gtmdsoft} in treating the TMD operator in (\ref{medef}). Physically,
the staple-shaped gauge link path incorporates the effect of final state
interactions on the struck quark in a semi-inclusive deep inelastic
scattering (SIDIS) process. The staple legs represent semi-classical quark
paths along which gluon exchanges with the proton remnant are
summed\footnote{To be specific, this interpretation pertains to the forward
in time, $\eta \geq 0$ branch in the convention adopted below, where $v$
points in the direction opposite to the proton momentum. Note that the quark
orbital angular momentum is an even function of $\eta $.}. Generalized
to the impact parameter-dependent case, the staple-shaped gauge link
path thus incorporates into the quark orbital angular momentum the torque
accumulated by the struck quark as it is leaving the proton \cite{burk}.
For $\eta =0$, this torque vanishes.

From the formal point of view, the TMD operator contains divergences that
are absorbed into renormalization and soft factors in the standard TMD
framework; these factors appear multiplicatively in the continuum theory
\cite{collbook,aybat,gtmdsoft,spl2}. They are
identical for all four instances of $\Phi $ in (\ref{ratiodef}) and the ratio
is therefore designed to cancel them (of course, these factors do not
depend on $\Delta_{T} $, which only enters through the external states).
This is the chief motivation for forming the ratio (\ref{ratiodef}) and
employing the nonlocal operator with separation $a$ in the denominator; in
this way, operators in the numerator and denominator match already at finite
lattice spacing. Setting the number of valence quarks $n$ to the appropriate
integer serves as the renormalization condition. Nonetheless, since the
operator separations in (\ref{ratiodef}) are proportional to the lattice
spacing, additional ultraviolet divergences arise as the lattice spacing
goes to zero. This is equivalent to the observation in momentum space that,
even if one has constructed a renormalized TMD, $k_T $-moments thereof
may still diverge as one lets $z_T $, which acts as a regulator on
$k_T $-integrals, go to zero. The specific scheme to control that
divergence adopted in (\ref{ratiodef}) amounts to identifying the
transverse momentum cutoff $z_T $ with the lattice resolution $a$.
To connect (\ref{ratiodef}) with its counterpart in other renormalization
schemes such as the standard $\overline{MS} $ scheme, an additional
matching would be required that is not available at present. The numerical
results presented below suggest that this unquantified systematic
uncertainty is minor. The scale evolution of quark orbital angular
momentum has been discussed in detail recently in \cite{hatta_evol},
giving an estimate of the variation expected in the regime in which
the lattice calculation is performed. The numerical calculation to
follow is carried out at a single lattice spacing $a$; it would be
interesting to extend it to several lattice spacings to directly
observe the scale evolution of the results.

It should be noted that the multiplicative nature of the renormalization
factors in the continuum theory does not straightforwardly extend to
the lattice theory. In general, the breaking of chiral symmetry
engendered by the Wilson fermion discretization used in this work
generates operator mixing within the class of TMD operators
\cite{rentmd,pertmix,auxmix,npmix} that precludes a simple factoring out
of renormalization factors and cancellation in the ratio (\ref{ratiodef}).
Also these effects will not be studied quantitatively in the present
work, and they constitute a further systematic uncertainty. A study
of the Sivers shift ratio \cite{rentmd}, in which the same TMD operator
appears as in (\ref{medef}), revealed no significant operator mixing effects
at the level of statistical accuracy achieved in the calculation; that
study included the gauge ensemble used also in the present work. This
suggests that operator mixing effects also do not introduce a dominant
systematic bias in the results obtained here.

A standard way to regulate the rapidity divergences \cite{rapidrev} contained
in the TMD operator for light-like staple direction $v$ is to take $v$ off
the light cone into the spacelike region \cite{collbook,aybat}, while
maintaining a zero transverse component, $v_T =0$. A convenient
Lorentz-invariant way to characterize the direction of $v$ is the
Collins-Soper type parameter
\begin{equation}
\hat{\zeta } = \frac{v\cdot P}{\sqrt{|v^2 |} \sqrt{P^2 } }
\label{zetadef}
\end{equation}
on which the numerical results obtained below will also depend.
Ultimately, one is interested in their large-$\hat{\zeta } $ behavior;
this corresponds to $v$ approaching the light cone.
Choosing $v$ to be spacelike simultaneously facilitates a straightforward
connection to the standard Lattice QCD methodology for evaluating hadronic
matrix elements: Given that the temporal dimension in Lattice QCD is
Euclidean, serving to project out the hadronic ground state, the operators
of which one evaluates matrix elements cannot be extended in physical time.
However, once a spacelike vector $v$ is adopted, the problem at hand can be
boosted to a Lorentz frame in which $v$ is purely spatial, and thus the
entire TMD operator in (\ref{medef}) exists at a single time. The lattice
calculation can be performed in that frame. Maintaining $v_T =0$ in the
lattice frame, $v$ will point purely in the (negative) 3-direction,
$v\equiv -\vec{e}_{3} $. In that case, $\hat{\zeta } = P_3 /m$ (where $m$
is the proton mass). Consequently, achieving large $\hat{\zeta } $ requires
large proton momentum component $P_3 $; this represents a significant
challenge, cf.~\cite{bmlat} for a study in the context of the Boer-Mulders
effect. On the other hand, in the special case $\eta =0$, the dependence on
the staple direction $v$ disappears; Ji quark orbital angular
momentum is boost-invariant. The corresponding results obtained
below will indeed be seen to be independent of $P_3 $ (or, equivalently,
$\hat{\zeta } $, if one formally maintains $v\equiv -\vec{e}_{3} $ in
the lattice frame).

On a lattice of finite extent with periodic boundary conditions in the
spatial directions, momenta are quantized. Therefore, performing direct
calculations only of the matrix element $\Phi (z_T )$, cf.~(\ref{medef}),
limits the accuracy in determining its derivative with respect to
$\Delta_{T} $, by forcing one to evaluate finite differences over, in
practice, substantial momentum increments. In the initial study \cite{jitojm}
of quark orbital momentum in the proton employing the GTMD approach laid
out here, this was the dominant source of systematic uncertainty.
It introduced a bias in the overall magnitude of the numerical data
approaching a factor 2. Although relative comparisons performed in
\cite{jitojm}, such as the one between Ji and Jaffe-Manohar orbital
angular momentum, can be expected to be robust with respect to this
bias, in absolute terms, the Ji orbital angular momentum extracted
in \cite{jitojm} displayed a significant discrepancy compared to the
value obtained independently via Ji's sum rule. To remedy this
dominant systematic bias, and thereby achieve agreement with the result
obtained via Ji's sum rule within the statistical fluctuations, is the
principal objective and advance of the present work. Other systematic
uncertainties, such as the ones associated with renormalization
discussed further above, are, in comparison, minor, and are
accordingly deferred to future work.

To completely remove the systematic bias originating from finite
difference evaluations of the $\Delta_{T} $-derivative of $\Phi (z_T )$,
a direct derivative method is adopted in the present work. The detailed
implementation of the method will be described in the next section,
following \cite{rome}, where the method was first laid out in detail.
Heuristically, the method is based on the observation that arbitrarily
small increments in overall momenta can be achieved by twisting the
spatial boundary conditions of the quarks, or, equivalently, coupling
the quarks to a constant $U(1)$ background gauge field. For the purpose
of computing a momentum derivative, this gauge field can be infinitesimally
small, and can therefore be treated perturbatively. In effect, this
generates an additional vector current insertion in the diagram one
would calculate to obtain $\Phi (z_T )$. Thus, by instead directly
performing a lattice calculation of the diagram containing the additional
vector current insertion, one directly accesses the $\Delta_{T} $-derivative
of $\Phi (z_T )$, excluding any systematic bias. A moderate price one pays
is that the additional operator insertion will tend to somewhat increase
the statistical fluctuations in the calculation.

\section{Lattice methodology}
To obtain the proton matrix element $\Phi (z_T )$, cf.~(\ref{medef}), one
calculates three-point functions $C_{\mbox{\scriptsize 3pt} } [\hat{O} ]$
together with two-point functions $C_{\mbox{\scriptsize 2pt} } $, which are
projected onto a definite proton momentum
${\bf p^{\prime } } = {\bf P} +\boldsymbol{\Delta }_{T} /2$ at the proton
sink, as well as a definite momentum transfer $\boldsymbol{\Delta }_{T} $
at the operator insertion in $C_{\mbox{\scriptsize 3pt} } [\hat{O} ]$,
\begin{eqnarray}
C_{\mbox{\scriptsize 3pt} } [\hat{O} ] (t,t_f ,{\bf p^{\prime } }, {\bf p} )
&=& \sum_{{\bf x}_{f} , \, {\bf y} }
e^{-i{\bf x}_{f} \cdot {\bf p^{\prime } }
+i {\bf y} \cdot ({\bf p^{\prime } -p} ) }
\mbox{tr} [\Gamma_{\mbox{\scriptsize pol} }
\langle n (t_f ,{\bf x}_{f} ) \hat{O} (t,{\bf y} )
\bar{n} (0,0) \rangle ] \\
C_{\mbox{\scriptsize 2pt} } (t_f ,{\bf p^{\prime } } ) &=&
\sum_{{\bf x}_{f} }
e^{-i{\bf x}_{f} \cdot {\bf p^{\prime } } }
\mbox{tr} [\Gamma_{\mbox{\scriptsize pol} }
\langle n (t_f ,{\bf x}_{f} )
\bar{n} (0,0) \rangle ] \ .
\end{eqnarray}
By momentum conservation, a definite source momentum
${\bf p} ={\bf P} -\boldsymbol{\Delta }_{T} /2$
is thereby implied in $C_{\mbox{\scriptsize 3pt} } [\hat{O} ]$. The proton
interpolating fields $n(t,{\bf x} )$ are constructed in practice using
Wuppertal-smeared quarks, to be discussed in more detail below, such as
to optimize overlap with the true proton state. The projector
$\Gamma_{\mbox{\scriptsize pol} } =
\frac{1}{2} (1+\gamma_{4} ) \frac{1}{2} (1-i\gamma_{3} \gamma_{5} ) $
selects states polarized in the 3-direction. As already discussed
further above, the lattice calculation is performed in a Lorentz frame
in which the TMD operator $\hat{O} $, specified in (\ref{medef}), exists
at the single time $t$; in particular, $v=-\vec{e}_{3} $, corresponding
to $\hat{\zeta } =P_3 /m$. In general, the three-point function
$C_{\mbox{\scriptsize 3pt} } [\hat{O} ]$ contains both connected
contributions, in which $\hat{O} $ is inserted into a valence quark
propagator, as well as disconnected contributions, in which $\hat{O} $
is inserted into a sea quark loop. In the present investigation, only
the former are taken into account. The latter contributions, which are
associated with significantly higher computational cost, are expected
to be minor, and are excluded. In the isovector $u-d$ quark channel,
the disconnected contributions cancel exactly; in that case, no
systematic uncertainty is associated with neglecting the disconnected
diagrams.

Having calculated the three-point and two-point functions, the matrix
element (\ref{medef}) is obtained from the ratio
\begin{equation}
2 E({\bf p^{\prime } } )
\frac{C_{\mbox{\scriptsize 3pt} } [\hat{O} ]
(t,t_f ,{\bf p^{\prime } } ,{\bf p} )}{C_{\mbox{\scriptsize 2pt} }
(t_f ,{\bf p^{\prime } } )}
\longrightarrow \Phi (z_T )
\label{3to2rat}
\end{equation}
which exhibits plateaus in $t$ for $0 \ll t \ll t_f $, yielding $\Phi (z_T )$.
Here, $E({\bf p } )=E({\bf p^{\prime } } )$ denotes the energy of the initial
and final proton states. Note that, for general choices of initial and final
momenta, the ratio (\ref{3to2rat}) has to be replaced by a more
general expression \cite{LHPC_1,LHPC_2}; it is the specific symmetric
choice of the initial and final momenta
${\bf p} ={\bf P} -\boldsymbol{\Delta }_{T} /2$,
${\bf p^{\prime } } ={\bf P} +\boldsymbol{\Delta }_{T} /2 $
that allows one to use the simple
ratio (\ref{3to2rat}) here. For finite temporal separations, residual
excited state contributions will contaminate the extraction of plateaus
from (\ref{3to2rat}). Control over these can be obtained by employing
a sequence of source-sink separations $t_f $. In the present study,
data for only one source-sink separation $t_f =1.14\, \mbox{fm} $
were gathered, and therefore it will not be possible to estimate excited
state effects quantitatively. Previous form factor studies on the same
gauge ensemble \cite{lhpc_ff1,lhpc_ff2} showed that the importance of
excited state effects depends substantially on the specific observable
considered. In some cases, the systematic bias at the separation
$t_f =1.14\, \mbox{fm} $ was seen to be smaller than the statistical
fluctuations, in other cases, it exceeded the latter by factors up to
2 to 3. Controlling for excited state effects will certainly be desirable
in future investigations.

Consider now evaluating the $\Delta_{T} $-derivative of $\Phi (z_T ) $
at $\Delta_{T} =0$, as called for by (\ref{ratiodef}). First, note that the
two-point function $C_{\mbox{\scriptsize 2pt} } $ is an even function of 
$\boldsymbol{\Delta }_{T} $; therefore, only the derivative of the three-point
function $C_{\mbox{\scriptsize 3pt} } [\hat{O} ]$ is needed\footnote{In
applications requiring higher derivatives with respect to momentum
transfer, such as calculations of charge radii \cite{nhasan}, also
derivatives of the two-point function enter.}, while one can directly
use $C_{\mbox{\scriptsize 2pt} } (t_f ,{\bf P} ) $ in (\ref{3to2rat}).
To proceed, it is useful to write $C_{\mbox{\scriptsize 3pt} } [\hat{O} ]$
explicitly in terms of the appropriate propagators, combining the
coordinates for ease of notation into four-vectors, e.g.,
$(t,{\bf y} ) \equiv y$,
\begin{eqnarray}
& & C_{\mbox{\scriptsize 3pt} } [\hat{O} ]
(t,t_f ,{\bf p^{\prime } }, {\bf p} )
=\sum_{{\bf x}_{f} , \, {\bf y} }
e^{-i{\bf x}_{f} \cdot {\bf P} +
i({\bf y} -{\bf x}_{f} /2) \cdot \boldsymbol{\Delta }_{T} } \cdot \\
& & \hspace{0.9cm} \cdot \left\langle \mbox{tr} \left[
S_{\Gamma_{\mbox{\tiny pol} } }^{n\bar{n} } (0;x_f )
G_{\mbox{\scriptsize sm-pt} } (x_f , y-z_T /2 ) \gamma^{+} U(y-z_T /2,y+z_T /2)
G_{\mbox{\scriptsize pt-sm} } (y+z_T /2,0) \right] \right\rangle \nonumber \\
& & \hspace{2.9cm}
=\sum_{{\bf x}_{f} , \, {\bf y} }
e^{-i{\bf x}_{f} \cdot {\bf P} } \ e^{-i({\bf x}_{f}
-({\bf y} -{\bf z}_{T} /2)) \cdot \boldsymbol{\Delta }_{T} /2} \
e^{i({\bf y} +{\bf z}_{T} /2) \cdot \boldsymbol{\Delta }_{T} /2} \cdot \\
& & \hspace{0.9cm} \cdot \left\langle \mbox{tr} \left[ \left(
\gamma_{5} G_{\mbox{\scriptsize pt-sm} } (y-z_T /2, x_f ) \gamma_{5}
S_{\Gamma_{\mbox{\tiny pol} } }^{n\bar{n} \, \dagger } (0;x_f )
\right)^{\dagger } \gamma^{+} U(y-z_T /2,y+z_T /2)
G_{\mbox{\scriptsize pt-sm} } (y+z_T /2,0) \right] \right\rangle \nonumber \\
& & \hspace{2.9cm}
=\sum_{{\bf x}_{f} , \, {\bf y} }
e^{-i{\bf x}_{f} \cdot {\bf P} } \cdot \\
& & \hspace{0.9cm} \cdot \left\langle \mbox{tr} \left[ \left( \gamma_{5}
G_{\mbox{\scriptsize pt-sm} } (y-z_T /2, x_f;\boldsymbol{\Delta }_{T} /2)
\gamma_{5} S_{\Gamma_{\mbox{\tiny pol} } }^{n\bar{n} \, \dagger } (0;x_f )
\right)^{\dagger } \gamma^{+} U(y-z_T /2,y+z_T /2)
G_{\mbox{\scriptsize pt-sm} } (y+z_T /2,0;-\boldsymbol{\Delta }_{T} /2)
\right] \right\rangle . \nonumber
\end{eqnarray}
Here, $S_{\Gamma_{\mbox{\tiny pol} } }^{n\bar{n} } $ denotes the standard
sequential source formed at the sink time $t_f $, as described in detail,
e.g., in the Appendix of \cite{dolgov}; $G_{\mbox{\scriptsize pt-sm} } (r,s)$
is the standard quark propagator from a smeared source to a point sink.
In the second step, the $\gamma_{5} $-hermiticity of the propagators was
used, and in the third step, the twisted propagator
\begin{equation}
G_{\mbox{\scriptsize pt-sm} } (r,s;{\bf q} ) =
e^{-i({\bf r} -{\bf s } ) \cdot {\bf q} }
G_{\mbox{\scriptsize pt-sm} } (r,s)
\end{equation}
was introduced. Thus, the $\Delta_{T} $-dependence has been entirely
absorbed into the twisted propagators, and it is their derivatives one
needs in order to obtain the $\Delta_{T} $-derivative of
$C_{\mbox{\scriptsize 3pt} } [\hat{O} ]$.
Following \cite{rome,nhasan}, in the absence of smearing, the derivative of
the twisted point-to-point propagator can be cast in the form
\begin{equation}
\left. \frac{\partial}{\partial q^j } 
G_{\mbox{\scriptsize pt-pt} } (r,s;{\bf q} )
\right|_{{\bf q} =0} = -i \sum_{z} G_{\mbox{\scriptsize pt-pt} } (r,z)
\Gamma_{V}^{j} G_{\mbox{\scriptsize pt-pt} } (z,s)
\end{equation}
where the sum extends over the four-dimensional coordinate $z$, and the
vector current insertion acts as
\begin{equation}
\Gamma_{V}^{j} G_{\mbox{\scriptsize pt-pt} } (z,s) =
U^{\dagger }_{j} (z-\vec{e}_{j} ) \frac{1+\gamma^{j} }{2}
G_{\mbox{\scriptsize pt-pt} } (z-\vec{e}_{j} ,s)
-U_j (z) \frac{1-\gamma^{j} }{2}
G_{\mbox{\scriptsize pt-pt} } (z+\vec{e}_{j} ,s) \ .
\end{equation}
Supplementing the point-to-point propagator with a smearing kernel,
\begin{equation}
G_{\mbox{\scriptsize pt-sm} } (r,s;{\bf q} ) =
e^{-i({\bf r} -{\bf s } ) \cdot {\bf q} }
\sum_{u} G_{\mbox{\scriptsize pt-pt} } (r,u) K(u,s)
=\sum_{u} G_{\mbox{\scriptsize pt-pt} } (r,u;{\bf q} ) K(u,s;{\bf q} )
\end{equation}
where also the twisted smearing kernel
\begin{equation}
K(u,s;{\bf q} ) = e^{-i({\bf u} -{\bf s } ) \cdot {\bf q} } K(u,s)
\end{equation}
has been introduced, one has the derivative
\begin{equation}
\left. \frac{\partial }{\partial q^j }
G_{\mbox{\scriptsize pt-sm} } (r,s;{\bf q} ) \right|_{{\bf q} =0} =
\sum_{z} G_{\mbox{\scriptsize pt-pt} } (r,z) \left[
-i \sum_{u} \Gamma_{V}^{j} G_{\mbox{\scriptsize pt-pt} } (z,u) K(u,s)
+\left. \frac{\partial }{\partial q^j }
K(z,s;{\bf q} ) \right|_{{\bf q} =0} \right]
\end{equation}
where the derivative of the twisted smearing kernel will be treated below.
Thus, in order to calculate the $\Delta_{T} $-derivative of
$C_{\mbox{\scriptsize 3pt} } [\hat{O} ]$, one has to evaluate two
additional propagators compared to a standard calculation of
$C_{\mbox{\scriptsize 3pt} } [\hat{O} ]$ itself. In the latter
case, one needs to evaluate the forward propagator from a smeared
source $K$, and the backward propagator from the smeared sequential
source $K\gamma_{5} S_{\Gamma_{\mbox{\tiny pol} } }^{n\bar{n} \dagger } $;
now, one additionally needs propagators from the sources
\begin{equation}
-i \Gamma_{V}^{j} G_{\mbox{\scriptsize pt-pt} } K
+\left. \frac{\partial }{\partial q^j } K \right|_{{\bf q} =0} \ ,
\ \ \ \ \ \ \ \ \ \ \ \ \ \
-i \Gamma_{V}^{j} G_{\mbox{\scriptsize pt-pt} }
K \gamma_{5} S_{\Gamma_{\mbox{\tiny pol} } }^{n\bar{n} \dagger }
+\left( \left. \frac{\partial }{\partial q^j } K \right|_{{\bf q} =0}
\right) \gamma_{5} S_{\Gamma_{\mbox{\tiny pol} } }^{n\bar{n} \dagger } \ .
\end{equation}
It remains to construct the derivative of the twisted smearing
kernel \cite{nhasan}. A single step of (twisted) Wuppertal smearing is
defined by
\begin{eqnarray}
K_0 (u,s;{\bf q} ) &=& e^{-i({\bf u} -{\bf s } ) \cdot {\bf q} }
\frac{1}{1+6\alpha } \left( \delta_{u,s} + \alpha \sum_{j=1}^{3}
\left[ U_j (u) \delta_{u+\vec{e}_{j} ,s}
+U^{\dagger }_{j} (u-\vec{e}_{j} ) \delta_{u-\vec{e}_{j} ,s} \right]
\right) \\
&=& \frac{1}{1+6\alpha } \left( \delta_{u,s} + \alpha \sum_{j=1}^{3}
\left[ e^{iq^j } U_j (u) \delta_{u+\vec{e}_{j} ,s}
+e^{-iq^j } U^{\dagger }_{j} (u-\vec{e}_{j} ) \delta_{u-\vec{e}_{j} ,s}
\right] \right)
\end{eqnarray}
so that its derivative at zero momentum is
\begin{equation}
K_0^{\prime } (u,s) \equiv
\left. \frac{\partial }{\partial q^j }
K_0 (u,s;{\bf q} ) \right|_{{\bf q} =0} =
\frac{\alpha }{1+6\alpha } \left[ i U_j (u) \delta_{u+\vec{e}_{j} ,s}
-i U^{\dagger }_{j} (u-\vec{e}_{j} ) \delta_{u-\vec{e}_{j} ,s} \right] \ .
\end{equation}
If the smearing kernel $K$ is given by $N$ steps of Wuppertal
smearing,
\begin{equation}
K (u,s;{\bf q} ) = \sum_{w_1 ,w_2 ,\ldots ,w_{N-1} }
K_0 (u,w_1 ;{\bf q} ) K_0 (w_1 ,w_2 ;{\bf q} ) \ldots
K_0 (w_{N-1} ,s;{\bf q} )
\end{equation}
then its derivative at zero momentum can be computed iteratively as
\begin{equation}
K^{\prime } \equiv (K_0^N )^{\prime } = K_0^{\prime } K_0^{N-1}
+K_0 (K_0^{N-1} )^{\prime } \ .
\end{equation}

The numerical data for the present study were generated using a 2+1-flavor
isotropic clover fermion ensemble on $32^3 \times 96$ lattices
generated by R.~Edwards, B.~Jo\'{o} and K.~Orginos
with lattice spacing $a=0.114\, \mbox{fm} $ and pion mass
$m_{\pi } =317\, \mbox{MeV} $. The present investigation is therefore
also significantly closer to the physical limit than the initial
exploration \cite{jitojm}. A total of 23224 data samples was gathered
on 968 gauge configurations. The Euclidean temporal separation between
proton sources and sinks was $t_f =10a$. HYP-smearing was applied to the
lattice links used in constructing the Wilson line $U$ in (\ref{medef}). This
leads to the renormalization and soft factors associated with the operator
in (\ref{medef}) corresponding more closely to their tree-level values even
before their cancellation in the ratio (\ref{ratiodef}). Three spatial proton
momenta were used, ${\bf P} \cdot L/(2\pi )=(0,0,n_P )$ with
$n_P =0,1,2$, where $L=32a$ denotes the spatial lattice extent.
Although the corresponding range of the Collins-Soper parameter
$\hat{\zeta } = 0,0.315,0.63$ is limited, and one cannot a priori
expect to obtain a good indication of the large-$\hat{\zeta } $ behavior
with data in this range, it will be seen below that the results for
Jaffe-Manohar orbital angular momentum already appear to stabilize
in the region of the two nonzero values of $\hat{\zeta } $. Corroboration
concerning this suggested early onset of asymptotic behavior by future
studies including larger $\hat{\zeta } $ would certainly be desirable.

\section{Numerical results}
\begin{figure}
\centerline{\psfig{file=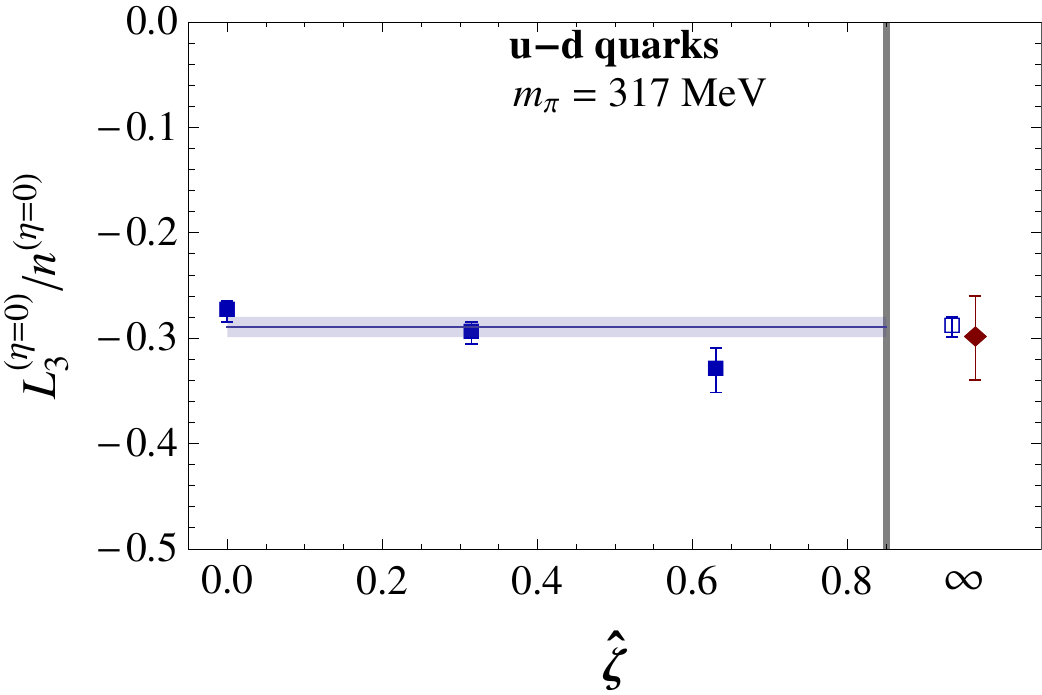,width=8.5cm} }
\caption{
Ji quark orbital angular momentum, i.e., the $\eta =0$ limit, for the
three values of $\hat{\zeta } $ probed, with the average plotted at
$\hat{\zeta } =\infty $ (open square). The filled diamond represents
the value extracted at the same pion mass in the $\overline{MS} $ scheme at
the scale $\mu^{2} = 4\, \mbox{GeV}^{2} $ via Ji's sum rule \cite{LHPC_2}.
The isovector $u-d$ quark combination was evaluated. The shown
uncertainties are statistical jackknife errors.}
\label{jiplot}
\end{figure}

Considering initially the special case of a straight Wilson line,
$\eta =0$, corresponding to Ji quark orbital angular momentum,
Fig.~\ref{jiplot} displays the results obtained in the isovector
case at the three available values of $\hat{\zeta } $. Recall that
$\hat{\zeta } $ is defined here through $v=-\vec{e}_{3} $ in the
lattice frame even when $\eta =0$; with that definition, on the other
hand, Ji quark orbital angular momentum should then be independent of
$\hat{\zeta } $, since $v$ does not in fact enter its construction.
This is borne out by the data in Fig.~\ref{jiplot}. The residual
apparent trend in the data may be due to the deviation between the
lattice dispersion relation and the continuum dispersion relation
used in the data analysis, but is also consistent with statistical
fluctuation.

Performing a $\chi^{2} $ fit of a constant in $\hat{\zeta } $ to the data
yields the average value plotted at $\hat{\zeta } =\infty $, taken here
to label the physical limit. This result is confronted with an
independent lattice determination of Ji quark orbital angular momentum via
Ji's sum rule at the same pion mass, in the $\overline{MS} $ scheme 
at the scale $\mu^{2} = 4\, \mbox{GeV}^{2} $ \cite{LHPC_2}.
The two determinations are in good agreement; the discrepancy observed in
the initial exploration \cite{jitojm} is entirely resolved. This validates
the use of the GTMD approach, properly implemented in particular with
respect to taking the $\Delta_{T} $-derivative in (\ref{ratiodef}), to
calculating quark orbital angular momentum in the proton. The result
corroborates the assumption that the various further systematic uncertainties
noted above, i.e., stemming from renormalization and matching, operator
mixing, or excited state contaminations are minor and do not bias the
result beyond the statistical uncertainty.

\begin{figure}
\centerline{\psfig{file=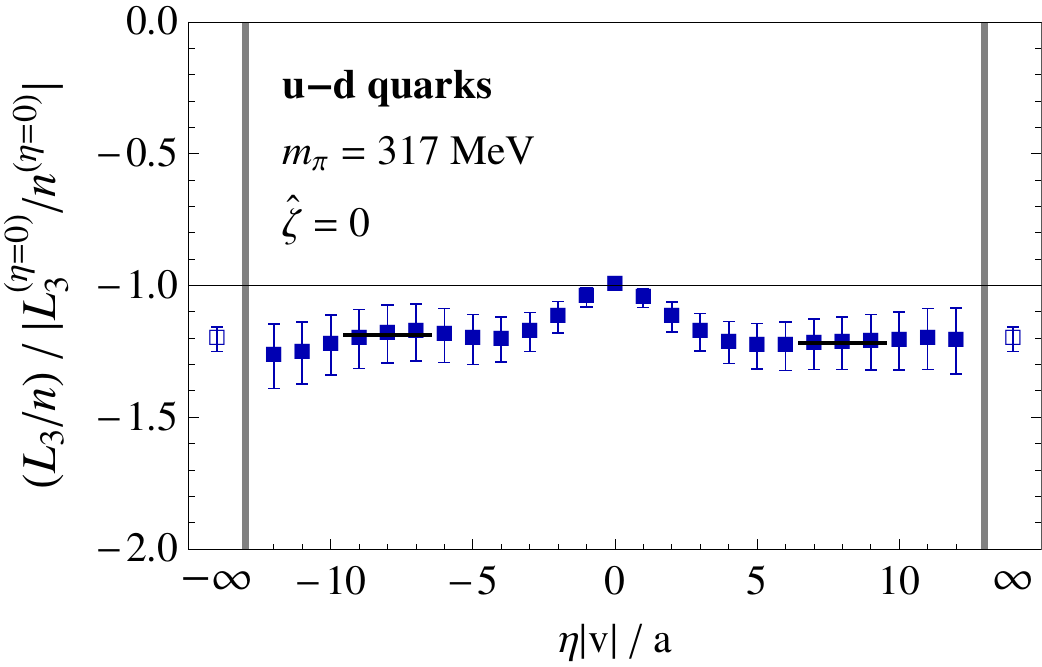,width=8.5cm} }
\centerline{\psfig{file=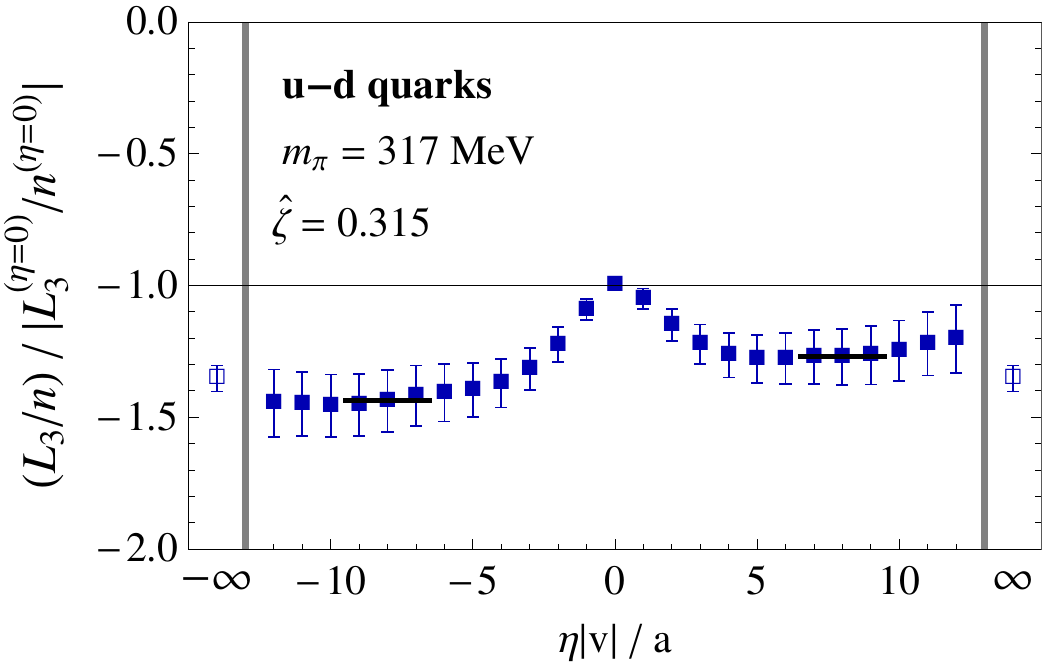,width=8.5cm} }
\centerline{\psfig{file=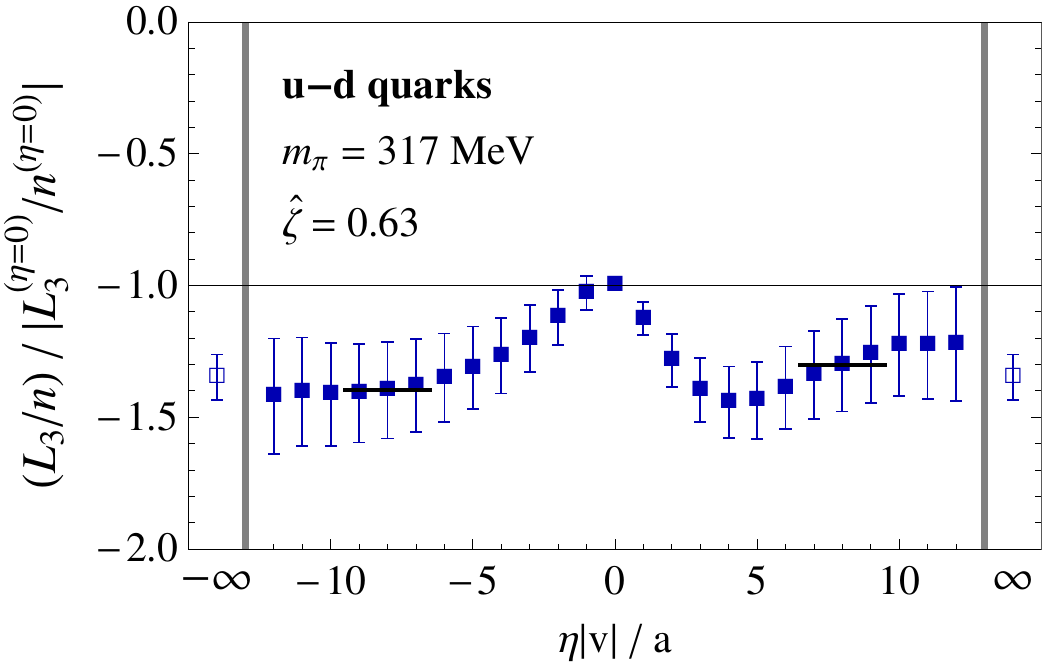,width=8.5cm} }
\caption{
Quark orbital angular momentum as a function of staple length parameter
$\eta $, normalized to the magnitude of Ji quark orbital angular momentum,
i.e., the result obtained at $\eta =0$. This quantity is even under
$\eta \rightarrow -\eta $, corresponding to time reversal. Accordingly, the
$|\eta |\rightarrow \infty $ extrapolated values are obtained by
averaging the $\eta >0$ and $\eta <0$ plateaus, which are determined
by fitting to the $|\eta | |v|/a =7,\ldots ,9$ range. The isovector
$u-d$ quark combination is shown, with the three panels corresponding
to the three available values of $\hat{\zeta } $. The shown uncertainties
are statistical jackknife errors.}
\label{etaplot}
\end{figure}

Departing from the $\eta =0$ limit, one probes the torque \cite{burk}
due to final state interactions accumulated by a quark struck in a deep
inelastic scattering process as it exits the proton. The
$\eta =\pm \infty $ limit corresponds to Jaffe-Manohar orbital angular
momentum. By varying $\eta $ gradually, a gauge-invariant, continuous
interpolation between the Ji and Jaffe-Manohar limits can be exhibited.
This is shown in Fig.~\ref{etaplot}, again for the isovector $u-d$ quark
channel, and for the three values of $\hat{\zeta } $ probed. Note that
the plots are normalized to the magnitude of the Ji quark orbital angular
momentum, i.e., the result obtained at $\eta =0$.

\begin{figure}
\centerline{\psfig{file=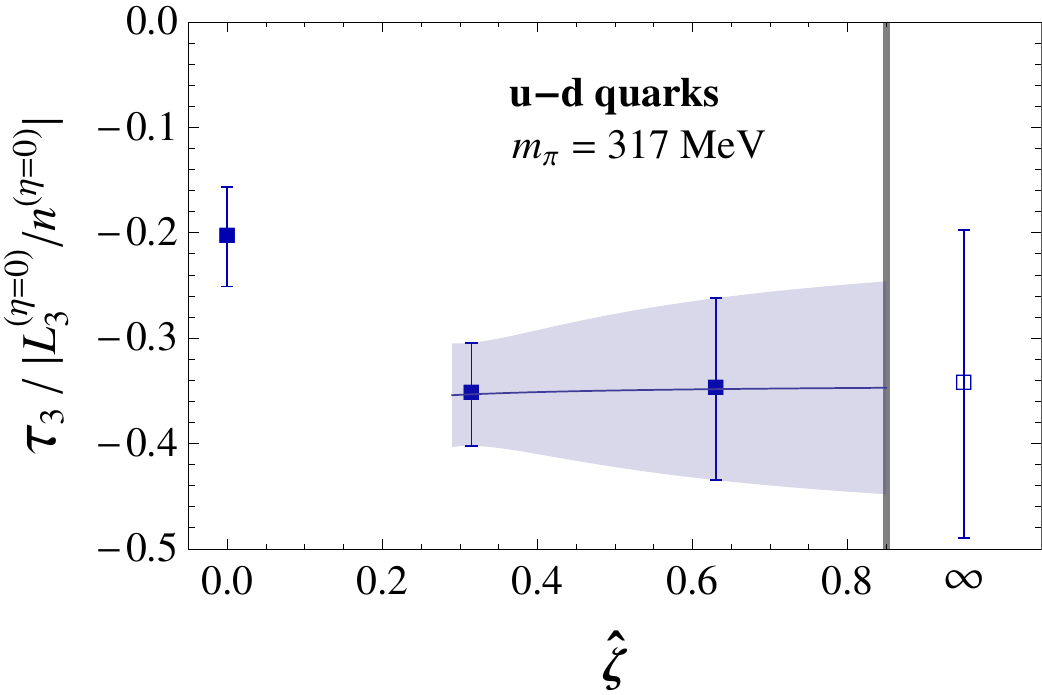,width=8.5cm} }
\caption{
Integrated torque accumulated by a quark struck in a deep inelastic
scattering process along its trajectory exiting the proton, as a
function of Collins-Soper parameter $\hat{\zeta } $. The data pertain
to the isovector $u-d$ quark channel, and are normalized to the
magnitude of the $\eta =0$ Ji orbital angular momentum. An ad hoc
extrapolation to the $\hat{\zeta } \rightarrow \infty $ limit is
also exhibited. The shown uncertainties are statistical jackknife errors.}
\label{torqueplot}
\end{figure}

Evidently, the torque supplied by the final state interactions is
appreciable, as already observed in \cite{jitojm}. Compared to the initial
Ji value, quark orbital angular momentum is enhanced in magnitude as one
proceeds towards the asymptotic Jaffe-Manohar limit. Contrasting the
three panels in Fig.~\ref{etaplot}, the effect strengthens as the
Collins-Soper parameter $\hat{\zeta } $ is increased, with Jaffe-Manohar
quark orbital angular momentum enhanced by about 30\% relative to the
Ji case for the two nonzero values of $\hat{\zeta } $. This is somewhat
less strong than in the exploration \cite{jitojm}; whether this is a genuine
physical trend associated with the change in pion mass from
$518\, \mbox{MeV} $ to $317\, \mbox{MeV} $, or whether it is an
artefact of the systematic bias in the calculation in \cite{jitojm}
cannot be decided at this point. The fact that the effect strengthens
with rising $\hat{\zeta } $ suggests that it can be expected to persist
into the $\hat{\zeta } \rightarrow \infty $ limit. Fig.~\ref{torqueplot}
displays the integrated torque, i.e., the difference between the
Jaffe-Manohar and Ji quark orbital momenta,
\begin{equation}
\tau_{3} = \frac{L^{(\eta = \infty )}_{3} }{n^{(\eta = \infty )} }
-\frac{L^{(\eta = 0)}_{3} }{n^{(\eta = 0)} } \ ,
\label{torquedef}
\end{equation}
as a function of $\hat{\zeta } $, normalized to the magnitude of Ji
quark orbital momentum. An extrapolation to the
$\hat{\zeta } \rightarrow \infty $ limit is also shown.
The ad hoc fit ansatz, $A+B/\hat{\zeta } $, is not underpinned by a
theoretical argument at this point, but is motivated by the good
description it provides of the considerably more detailed data as a
function of $\hat{\zeta } $ available for the pion Boer-Mulders TMD
ratio \cite{bmlat}. Auxiliary information concerning the expected
large-$\hat{\zeta } $ behavior would be desirable to aid in
sharpening the analysis. The ad hoc extrapolation indeed yields a
signal in the $\hat{\zeta } \rightarrow \infty $ limit.

\begin{figure}
\centerline{\psfig{file=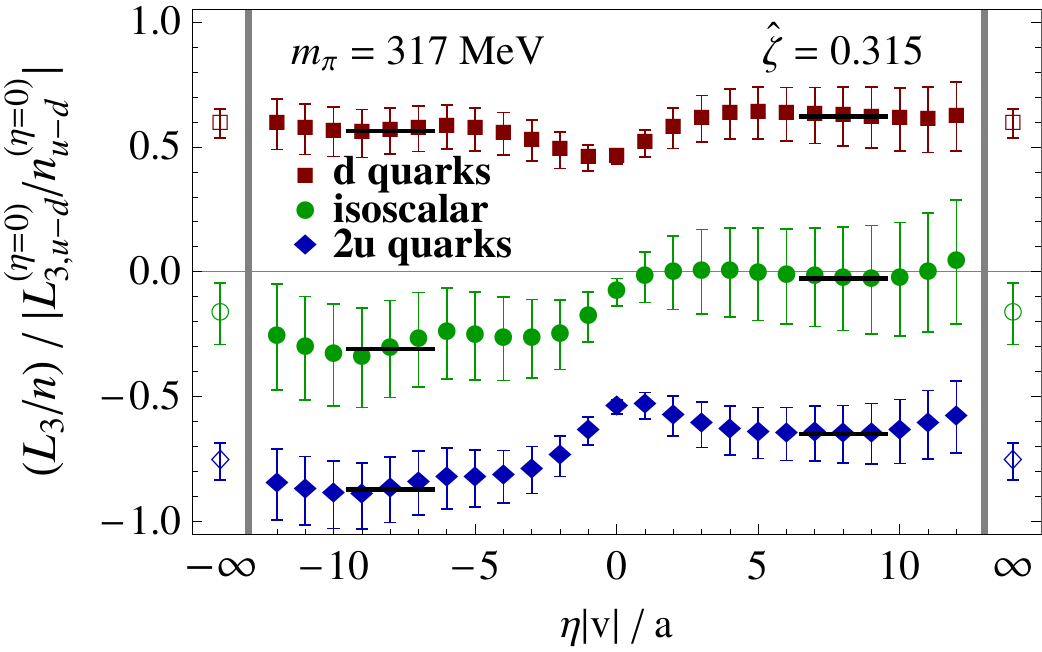,width=8.5cm} }
\caption{
Flavor-separated quark orbital angular momentum as a function of staple
length $\eta $, analogous to Fig.~\ref{etaplot}, at fixed
$\hat{\zeta } =0.315$. Results are displayed for $d$ quarks and for two $u$
quarks (i.e., the $u$-quark data for $L_3 /n$ have been multiplied by
$2$ to compensate for $n=2$), as well as for the isoscalar total
quark orbital angular momentum, i.e., the sum of the ``$d$'' and
``$2u$'' data. All results are still normalized by the magnitude of
isovector Ji orbital angular momentum (thus, at $\eta =0$, the ``$2u$''
and ``$d$'' data differ by unity). The shown uncertainties are statistical
jackknife errors.}
\label{flav_etaplot}
\end{figure}

Generalizing to the flavor-separated case, it should be kept in mind
that the additional disconnected contributions that arise compared
to the isovector case have not been evaluated. These are, however,
expected to be minor at the pion mass $m_{\pi } = 317\, \mbox{MeV} $
used in this calculation. Fig.~\ref{flav_etaplot} shows data analogous
to Fig.~\ref{etaplot} for one value of $\hat{\zeta } $, exhibiting the
behavior of $d$-quark and $u$-quark orbital angular momentum separately,
as well as the total (isoscalar) quark orbital angular momentum. Here,
the $u$-quark data have been normalized to two quarks, i.e.,
$L_3 /n$ in the $u$ quark case has been multiplied by 2 to compensate
for $n=2$ for $u$ quarks; hence the ``$2u$'' label. The isoscalar result
was then obtained by simple addition of the ``$d$'' and ``$2u$''
data\footnote{This may differ slightly from calculating
$3L_{3,u+d} /n_{u+d} $ at finite statistics.}.

\begin{figure}
\centerline{\psfig{file=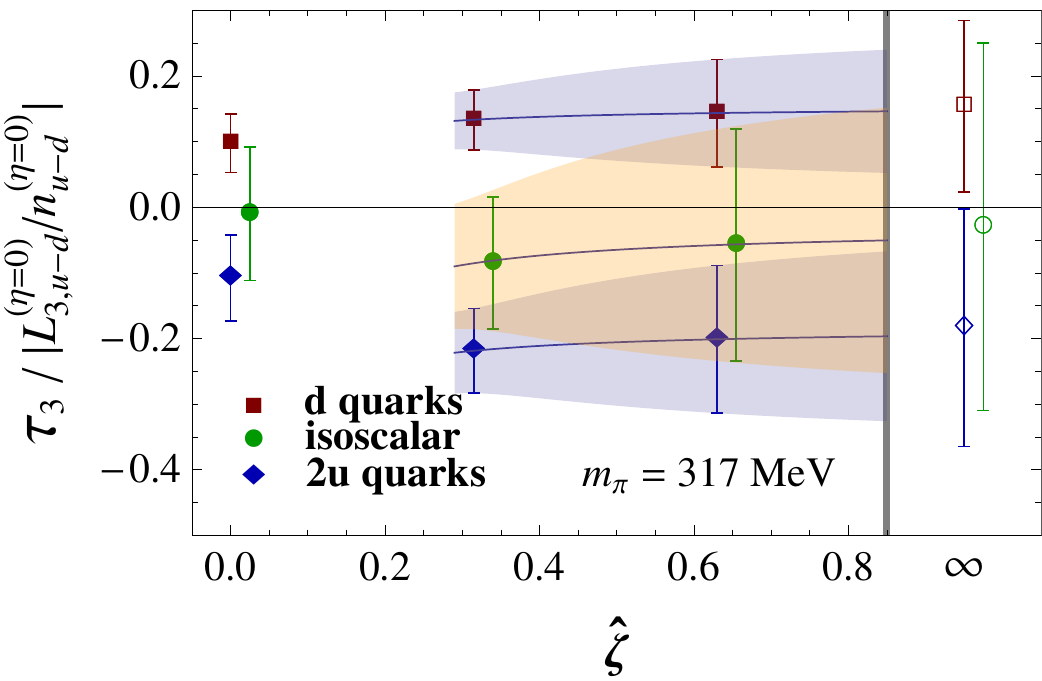,width=8.5cm} }
\caption{Flavor-separated integrated torque accumulated by a quark struck
in a deep inelastic scattering process, as a function of $\hat{\zeta } $,
analogous to Fig.~\ref{torqueplot}, together with ad hoc extrapolations
to infinite $\hat{\zeta } $. Data for $d$ quarks, for two
$u$ quarks (i.e., the $u$-quark data for $\tau_{3} $ have been
multiplied by $2$ to compensate for $n=2$ in the $u$-quark case),
and their sum, the isoscalar $u+d$ quark combination, are shown.
Data are normalized by the magnitude of isovector Ji quark orbital angular
momentum, as in previous figures. For better visibility, isoscalar
data are slightly displaced horizontally. The shown uncertainties
are statistical jackknife errors.}
\label{flav_torqueplot}
\end{figure}

As observed previously in \cite{jitojm}, the strong cancellation of the
$u$- and $d$-quark orbital angular momenta in the proton that has
long been known for the $\eta =0$ Ji case \cite{LHPC_1,LHPC_2}
extends to nonzero $\eta $ and the Jaffe-Manohar limit. Only a
small negative contribution to the spin of the proton from quark
orbital angular momentum remains. The data for the flavor-separated
integrated torque, cf.~(\ref{torquedef}), are collected in
Fig.~\ref{flav_torqueplot} and extrapolated to the
$\hat{\zeta } \rightarrow \infty $ limit. At the present level
of statistics, scarcely a signal is obtained for the flavor-separated
integrated torque in that limit; the data do appear compatible
with the observations made at fixed $\hat{\zeta } =0.315$ from
Fig.~\ref{flav_etaplot}.

\section{Conclusions and outlook}
The chief advance of the present study is the adoption of a direct
derivative method \cite{rome} in the GTMD approach to evaluating quark
orbital angular momentum in the proton in Lattice QCD. The introduction of
this method has led to a reliable quantitative computation of the needed
derivative, with respect to momentum transfer, of the relevant GTMD
matrix element. This is validated by the result obtained specifically
for the quark orbital angular momentum defined through the Ji
decomposition of proton spin; it agrees well with the corresponding
result obtained independently in Lattice QCD calculations relying on
Ji's sum rule \cite{LHPC_2}. The discrepancy observed in the initial
exploration \cite{jitojm} is thus resolved.

The agreement between the quark orbital angular momentum calculated in
this work using the GTMD approach and the result from the Ji sum rule
suggests that other systematic uncertainties, such as ones associated
with excited state effects, renormalization and matching, as well as
operator mixing are minor and do not rise to the level of the statistical
uncertainties of the present calculation.

In the GTMD approach, one directly computes quark orbital angular
momentum by weighting the appropriate Wigner function (related to
GTMDs via Fourier transformation) by $b_T \times k_T $, where $b_T $
is the quark impact parameter and $k_T $ the quark transverse momentum
\cite{lorce,leader}.  The aforementioned derivative with respect to
momentum transfer supplies the weighting by $b_T $, its Fourier conjugate.
The information on $k_T $, on the other hand, is supplied through the
nonlocal TMD operator used in constructing the relevant proton matrix
element (\ref{medef}). The treatment of renormalization issues thus
follows closely the methods used in more widely explored Lattice TMD
calculations \cite{straightlett,straightlinks,tmdlat,bmlat,rentmd}.
Ratios of proton matrix elements are constructed to cancel
renormalization and soft factors associated with the TMD operator. In
effect, one evaluates quark orbital angular momentum in units of the
number of valence quarks. Operator mixing effects
\cite{rentmd,pertmix,auxmix,npmix} can spoil these cancellations,
and though they appear not to play a significant role in the present
calculation, these effects will ultimately have to be brought under
control.

The advantage of this nonlocal operator-based GTMD approach is that
one can extend Lattice QCD calculations beyond the Ji decomposition
of proton spin and establish a continuous, gauge-invariant interpolation
from Ji \cite{jidecomp} to Jaffe-Manohar \cite{jmdecomp} quark orbital
angular momentum. The corresponding information is contained in the choice
of Wilson line path in the TMD operator. A straight Wilson line path yields
Ji quark orbital angular momentum \cite{jist,burk,eomlir}, and a
staple-shaped Wilson line path, in the limit of infinite staple length,
yields its Jaffe-Manohar counterpart \cite{hatta,burk}. The difference
between the two can be interpreted as the integrated torque accumulated
by a quark struck in a deep inelastic scattering process as it exits the
proton, through final state interactions \cite{burk}. It corresponds to
a Qiu-Sterman type correlator \cite{burk,eomlir,eomlong}. The data obtained
for this term in the present investigation allow one to observe the
gradual accumulation of torque by the quark until it attains the
asymptotic Jaffe-Manohar orbital angular momentum. The latter is
enhanced in magnitude relative to the initial Ji value, with the
integrated torque adding about one third of the magnitude of the Ji
orbital angular momentum at the pion mass $m_{\pi } = 317\, \mbox{MeV} $
used in this calculation.

To further sharpen the analysis of quark orbital angular momentum
within the GTMD approach, calculations for a sequence of lattice
spacings would be desirable, to allow for a direct study of the
scale evolution. Also, a more comprehensive exploration of the
dependence on the Collins-Soper parameter $\hat{\zeta } $ is
warranted, to clarify whether the results indeed already stabilize at
the fairly low values of $\hat{\zeta } $ employed in this work, as
suggested by Fig.~\ref{torqueplot}. Ultimately, the large $\hat{\zeta } $
behavior determines the physical limit. Finally, of course, also further
progress towards the physical pion mass must be made.

The improved calculation of quark orbital angular momentum based on GTMDs
achieved using the present methodology opens the way to reliably compute
other related quantities, such as quark spin-orbit correlations in the
proton \cite{lorce,eomlong}. Corresponding calculations are in progress,
employing domain wall fermions to curtail the operator mixing effects that
are induced when the fermion discretization breaks chiral symmetry.

\section*{Acknowledgments}
This work benefited from fruitful discussions with M.~Burkardt, W.~Detmold,
R.~Gupta, S.~Liuti and C.~Lorc\'{e}.
The lattice calculations performed in this work
relied on code developed by B.~Musch, as well as the Chroma \cite{chroma}
and Qlua \cite{qlua} software suites.
R.~Edwards, B.~Jo\'{o} and K.~Orginos provided the clover fermion ensemble,
which was generated using resources provided by XSEDE (supported by 
National Science Foundation Grant No.~ACI-1053575). Computations were
performed using resources of the National Energy Research Scientific
Computing Center (NERSC), a U.S.~DOE Office of Science User Facility
operated under Contract No.~DE-AC02-05CH11231. This work was furthermore
supported by the U.S.~Department of Energy, Office of Science, Office
of Nuclear Physics under grant DE-FG02-96ER40965 (M.E.), grant DE-SC-0011090
(J.N.), grant DE-SC0018121 (A.P.), as well as through the TMD Topical
Collaboration (M.E., J.N.); and it was also supported by the
U.S.~Department of Energy, Office of Science, Office of High Energy Physics
under Award Number DE-SC0009913 (S.M.). N.H.~and S.K.~received support
from Deutsche Forschungsgemeinschaft through grant SFB-TRR 55, and
S.S.~is supported by the U.S.~National Science Foundation under CAREER
Award PHY-1847893 and through the RHIC Physics Fellow Program of the
RIKEN BNL Research Center.

\end{document}